\documentclass[prl,twocolumn,groupedaddress,showpacs,byrevtex,endfloats]{revtex4}

\begin{document}

\title{Role of bulk and surface phonons in the decay of metal surface states} 

\author{A. Eiguren$^{1}$, B. Hellsing$^{2}$, F. Reinert$^{3}$,
  G. Nicolay$^{3}$, E. V. Chulkov$^{1,4}$, V. M. Silkin$^{4}$,
  S. H\"{u}fner$^3$ and P. M. Echenique$^{1,4}$ }  
\vspace{2cm}
\address{1. Departmento de F\'{i}sica de Materiales and Centro Mixto
  CSIC-UPV/EHU, Facultad de Ciencias Qu\'{i}micas,  
Universidad del Pais Vasco/Euskal Herriko Unibertsitatea, Adpo. 1070,
20018 San Sebasti\'{a}n/Donostia, Basque Country, Spain} 

\address{2. Department of Physics, Chalmers University of Technology
and G\"{o}teborg University, 
S-412 96 G\"{o}teborg, Sweden}
\address{3. Fachrichtung Experimentalphysik, Universit\"{a}t des
  Saarlandes, 66041 Saarbr\"{u}cken, Germany}  
\address{4. Donostia International Physics Center, 
Paseo de Manuel Lardizabal, 4, 20018 San Sebasti\'{a}n/Donostia, Spain}

\date{\today}


\begin{abstract}

We present a comprehensive theoretical investigation of
the electron-phonon contribution to the lifetime broadening 
of the surface states on Cu(111) and Ag(111), in 
comparison with high-resolution photoemission results. The 
calculations, including electron and phonon states of 
the bulk and the surface, resolve the relative importance 
of the Rayleigh mode, being dominant for the lifetime 
at small hole binding energies. Including the 
electron-electron interaction, the theoretical results are in 
excellent agreement with the measured binding energy 
and temperature dependent lifetime broadening.
\end{abstract}


\pacs{73.20.At 79.60.Bm 71.38.-k}
%

\maketitle

Understanding the temporal evolution of quasi particles (electron and holes)
on metal surfaces is of paramount importance to describe many important phenomena 
such as the dynamics of charge and energy transfer, quantum interference, localization  
and many others. This temporal evolution is characterized by a finite lifetime, $\tau$, 
which refers to the time the quasi particle retains its identity. 
While the lifetime of an excited electron 
or hole is determined by many-body interactions, namely electron-electron
($e$-$e$) and electron-phonon ($e$-$p$) scattering processes, the
peak width in an experiment might also be influenced by 
electron-defect scattering on crystal or surface imperfections \cite{Theilmann97PRB}. 
However, it was demonstrated in recent STM \cite{Kliewer00S} and
photo\-emission experiments \cite{Reinert2001PRB}  
that these defect contributions can be minimized, making it possible to analyze the pure lifetime 
broadening due to the formation of a hole in the  $sp$ surface state
band in the $L$-gap of the (111)-surface of noble metals.

These Shockley-type surface states form a two-dimensional (2D) electron gas and the $e$-$e$ contribution 
to the hole lifetime has been rationalized in terms of a dominant contribution from 
intraband transitions within the 2D surface state band, screened by the underlying 
3D bulk electron system, and in terms of interband transitions (bulk 
states $\rightarrow$ surface state) \cite{Kliewer00S}. On the other hand an appropriate 
calculation of the  $e$-$p$ contribution to the lifetime broadening of surface states 
is still lacking. The present work is an attempt in this direction.

The strength of the  $e$-$p$ coupling is described by the electron 
mass enhancement parameter $\lambda$, which is, in general, energy and momentum 
dependent. Many properties of metals \cite{Grimvall81}, such as resistivity, 
specific heat and superconductivity, reflect the  $e$-$p$ coupling and can 
be expressed in terms of the Fermi surface-averaged $\lambda$-value. 
It also reflects the high temperature behavior of the broadening 
$\Gamma_{ep}=2 \pi \lambda k_{B} T$, and the  $e$-$p$ contribution to 
the renormalization of the mass $m^{*}=m(1 +\lambda)$.
The anisotropy 
of $\lambda$ is well known \cite{Khan82PRB} and is revealed in e.g. cyclotron resonance 
measurements \cite{Lee70PRB}.

Typically, the phonon contribution to the decay of surface states is estimated 
using the Debye phonon model. Within this model the Eliashberg spectral function of the $e$-$p$
interaction is proportional to the quadratic density of phonon states
$\alpha^{2}F(\omega)=\lambda ({\omega/\omega_{D}})^{2}$, where
$\omega_{D}$ is the Debye energy, $\lambda$ is usually obtained
from measurements or theoretical calculations of bulk properties
\cite{Grimvall81}. However, it is not obvious that this approach
should be adequate for surface state electrons or holes when the surface
state itself, as well as the surface phonon modes, are not taken into
account and furthermore, the low temperature $\Gamma_{ep}$ will depend on the model used. 
A more rigorous treatment of the $e$-$p$
contribution is needed  especially for surface states close to the
Fermi level, because for these states  the $e$-$e$ contribution is small
and the electron-phonon interaction becomes dominant even at low
temperatures. 

In this letter we present an analysis of the  $e$-$p$ coupling contribution 
to the lifetime broadening of surface electron states taking into account 
all electron and phonon states involved in the electron-phonon
scattering process, in comparison to new energy and temperature
dependent high-resolution photoemission data. The theoretical analysis
is based on a calculation of the full Eliashberg spectral 
function. With this approach we are able to resolve in details the
contributions from different phonon modes, in particular from the
Rayleigh surface mode and bulk phonons and the general temperature
dependence. We also obtain the high temperature behavior represented
by $\lambda$, which is given by the first reciprocal moment of the
Eliashberg function. We show that an approach based on (i)
Thomas-Fermi screened Ashcroft ion-electron potentials, (ii)
one-electron states with ${first \ principles}$ quality and (iii)
a simple model phonon calculation give results in good agreement
with recently published experimental data
\cite{Kliewer00S,Theilmann97PRB,Reinert2001PRB,McDougall95PRB} and
data presented here.  

The phonon induced lifetime broadening of a surface band state with momentum  
$\vec{k}_{i}$ and energy $\omega_{i}$ is given by 
\begin{eqnarray}
\Gamma_{ep}(\omega_{0}) = 2 \pi\int_{0}^{\omega_{m}} \{ \ \alpha^{2}F_{\vec{k}_{i}}(\omega) \
[ 1 + 2 \ n( \omega ) + \nonumber\\ f( \omega_{0} + \omega) \ 
- f(\omega_{0} - \omega) ] \ \} d \omega \ , \hfill\label{eq:G}
\end{eqnarray}
where $\omega_{m}$ is the maximum phonon frequency and
\begin{eqnarray}
\alpha^{2}F_{\vec{k}_{i}}(\omega) = 
\sum_{\vec{q},\nu,f}
|g^{\vec{q},\nu}_{i,f}|^{2} 
\delta(\omega -\omega_{\vec{q},\nu} ) 
\ \delta(\epsilon_{\vec{k}_{i}} - \epsilon_{\vec{k}_{f}}) , \hfill\label{eq:Elias}
\end{eqnarray}
where the sum is over final electron states $\vec{k}_{f}$ of band $f$
and phonon modes $\nu$ with momentum $\vec{q}$. 

In Eq. ( \ref{eq:Elias} ) we also make use of the quasi elastic 
scattering approximation. The $e$-$p$ coupling constant is given by
\begin{eqnarray}
\label{eq:g}
g^{\vec{q},\nu}_{i,f} = \sqrt{\frac{1}{ 2MN\omega_{\vec{q}\nu}}} \times
\langle f | \sum_{\mu}\vec{\sl{\epsilon}}_{\vec{q}\nu}(\vec{R}_{\mu}) \cdot \vec{\nabla}_{\vec{R}_{\mu}} \tilde{V}^{\mu}_{q}|i\rangle \ ,
\end{eqnarray}
summing over the layers $\mu$ of the slab. $M$ and $N$, are
the atomic mass and number of ions in each layer of the slab,
respectively, $\vec{\sl{\epsilon}}_{\vec{q},\nu}(\vec{R}_{\mu})$
denotes the complex phonon polarization vectors normalized over the
slab and $\tilde{V}^{\mu}_{q}$ is the screened ion-electron
potential. The electron states, of the form
\(\psi_{n,\vec{k}}(\vec{x},z) = \langle z|n \rangle
\exp(-\vec{k}\cdot\vec{x}) \) , 
where $\vec{k}$ and  $\vec{x}$ is the electron momentum and electron
coordinate parallel to the surface, respectively are derived from a slab 
model calculation. This model potential scheme by Chulkov et
al. \cite{Chulkov99SS,Chulkov98PRL,Echenique00CP}  produces wave
functions  
in good agreement with those obtained from {\it ab initio} calculations 
and energy spectra reproducing experimental data.  

The Umklapp processes can be neglected as the Fermi momentum of the
noble metal surface states are small ($<$ 0.12 a.u.) in comparison
with half the minimum reciprocal vector ($<$ 0.75 a.u.). Furthermore,
the maximum Fermi momentum vector of the final bulk states is $<$ 0.74
a.u., which implies that the final states are all confined to the first
Brillouin zone. Thus, the contribution from Umklapp processes is less
than 10$\%$ for small binding energies and vanishes completely
approaching the $\bar{\Gamma}$-point.  
To obtain the phonon dispersion relations and polarization vectors we
perform a calculation using the same force constant model as I of Black et
al. \cite{Black83SS}, 
but in addition, we expand the dynamic matrix in Gottleib polynomials \cite{TrullingerJMP} 
in order to optimize the representation of the surface phonon modes.
The force constant is fitted to
reproduce elastic constants and bulk phonon frequencies, with
surprisingly good agreement with He scattering experiments \cite{Harten85FDCS}.

The screened ion-electron potential is determined by the static dielectric 
function and the bare pseudo potential, \( \tilde{V}^{\mu}_{q}(z)  =  \ 
\int dz' \ \tilde{\epsilon}^{-1}(z,z';q) \ \tilde{V}^{\mu}_{bare}(z';q) \) , 
where $q$ is the modulus of the phonon momentum wave vector parallel
to the surface and $\tilde{V}^{\mu}_{bare}$ the $2D$-Fourier  
transform parallel to the surface of the bare ion-electron Ashcroft pseudo 
potentials \cite{Ashcroft66PR} . We have investigated the quality of the 
screening using the dielectric function according to Thomas-Fermi and 
within RPA (constructed from the eigen wave functions and eigen energies 
of a 31-slab calculation). The difference is about 1$\%$ for both 
$\lambda$ and $\Gamma_{ep}$ due to a compensating effect. Referring 
to the surface layer, Thomas Fermi screening is symmetric while RPA 
yields a screening slightly stronger below and slightly weaker above.

We want to point out that in almost all investigations of the
$e$-$p$ interaction, the relevant electron scattering takes place
close to the Fermi surface and only the mean $e$-$p$ coupling, averaged
over the Fermi surface, comes into play. When considering the lifetimes 
of surface states, the situation is quite different. The $e$-$p$ coupling 
becomes state dependent as the probed initial
electron state is fixed. For the surface states of the studied noble
metal surfaces Cu(111) and Ag(111), not only the Fermi surface
is of importance because the binding energy of the initial hole
state ranges from zero (Fermi level) to about 0.4 eV.

The intraband electron scattering ($f$=$i$) in Eq.(\ref{eq:g}) can be
neglected with the following argument. 
In the case of intraband transitions in the long wavelength limit,
the matrix element of the gradient of the screened ion-electron potential is approximately 
the expectation value of the force acting on the ions in the direction perpendicular 
to the surface
$\langle \psi_{i,\vec{k}}(\vec{x},z) \nabla_{\vec{R}_{\mu}} \tilde{V}^{\mu}(\vec{x},z) \psi_{i,\vec{k}-\vec{q}}(\vec{x},z) \rangle$ 
$\approx$ $\vec{F_{\mu}}$
$\approx$ $\langle i | \partial \tilde{V}^{\mu}_{0}(z)/ \partial z \ | i \rangle \hat{z}$.
It is well known that within linear response, the sum over the 
forces acting on the ions induced by the electron (hole) must be zero
\cite{Sorbello80PRB,Budd75PRB,Andersson84PRL,Persson87PS}, thus the matrix element of the 
sum of gradients of the screened ion-electron must be equal to zero $\sum_{\mu} \vec{F_{\mu}}$=$0$.
Furthermore, the ion displacements forming the phonon modes in the surface region, 
associated with small $q$ are locally rigid with a coherence length of $2\pi/q$ 
isotropically in all directions, in parallel and perpendicular to the
surface \cite{Flatte91PRB} 
From Eqs.\ (\ref{eq:G})---(\ref{eq:g}), it can be seen that this restriction and the 
reduced phase space in the intraband scattering process makes this contribution negligible 
compared to interband contribution. 

From Eq.\ \ref{eq:Elias} we see that the Eliashberg function is given
by the phonon density of states weighted by the electron-phonon
coupling strength $g$. In Fig 1. the calculated phonon dispersion and the
Eliashberg function is presented for Cu(111). The Rayleigh surface
mode is split of from the bulk phonon band, which gives a lower energy
peak in the Eliashberg function, at about $\sim$13 meV in Cu(111).
The oscillations in the Eliashberg function reflects the finite number
of layers of the model potential (31 layers) in the calculation of
electron wave functions and thus have no physical significance.   

The high-resolution photoemission experiments presented here were 
performed with an energy and angular resolution of $\Delta 
E\approx3$~meV and $\Delta \Theta\approx0.3^\circ$, respectively, at VUV 
photon energy of $h\nu=21.23$~eV (He~{\sc I}). A detailed description 
of experimental conditions and sample preparation can be found in 
Ref.\ \onlinecite{Reinert2001PRB}. A complete data set at one 
temperature, consisting of the photoemission data over an angular 
range of $\pm6^\circ$ off normal, was measured within the first 15~min 
after the last annealing step at $\approx600^\circ$C. As an example, the 
inset in Fig.~3 demonstrates the temperature dependence of the normal 
emission ($\bar{\Gamma}$ point) spectra on Cu(111) at three different 
temperatures. For the quantitative analysis we fitted the individual 
spectra of a temperature series as described in Ref.\ \onlinecite{Reinert2001PRB}. 
The resulting Lorentzian lifetime widths $\Gamma$ (full widths at half maximum: 
FWHM) are given as open symbols in Fig.~3. The binding energy dependence 
of the linewidth at low temperature (cf.\ Fig.~2) was determined from 
MDC (momentum distribution curve) cuts at the respective energy and are 
given relative to the linewidth at the Fermi level $E_F$, which is nearly 
completely determined by the finite angular resolution.

The most interesting hole binding energy region is of course very close 
to the Fermi level, in particular when the binding energy is less than the maximum 
phonon frequency $\omega_{m}$. In this region we expect that the 
lifetime broadening is completely determined by the the $e$-$p$ 
coupling. The contribution from $e$-$e$ interaction is very small, 
for Cu(111) and Ag(111),  $\Gamma_{ee}<0.2$ meV.

In Fig.~2 we present the calculated $\Gamma_{ep}$, at $T=30 K$, for
Cu(111) and Ag(111) together with the experimental results. From a
simple Debye model we would expect a cubic binding energy dependence
in the region below $\omega_{m}$, which obviously is not the case. The
saturation of $\Gamma_{ep}$ for binding energies exceeding
$\omega_{m}$ is clearly seen in experimental data. Adding the
contribution from the $e$-$e$ interaction, values close to the
experiment are obtained (see Tab.\ \ref{table}). We note from Fig. 2, that the
contribution from only the 
Rayleigh mode gives about 38$\%$ of $\Gamma_{ep}$ beyond the maximum
phonon frequencies, indicating that bulk phonons give most of the
contributions in this range. But for binding energies below the
maximum  of the Rayleigh mode energy, this mode alone represent on
average about the 85$\%$ of $\Gamma_{ep}$.

The main signature of the $e$-$p$ contribution to the lifetime
broadening is the temperature dependence. For binding energies
exceeding $k_{B}T$ we can neglect the temperature dependence of the
$e$-$e$ scattering. The temperature dependence of $\Gamma_{ep}$ was
calculated for the hole state in the $\bar{\Gamma}$ point for Cu(111)
and Ag(111). 
For a comparison with the experimental temperature dependence we show in Fig.~3 
the sum of the constant contribution from $\Gamma_{ee}$ and the $T$-dependent 
$\Gamma_{ep}$.
The calculated broadenings are in good agreement with the experimental data points, 
except for the highest temperature data point for Cu(111).

In the Tab. \ref{table} we summarize the calculated and measured low temperature
results for the lifetime broadening for the surface states of Cu(111)
and Ag(111). Summing up the calculated contribution from $e$-$e$
interaction $\Gamma_{ee}$ \cite{Kliewer00S} and the results of the
present work for the $e$-$p$ interaction $\Gamma_{ep}$ show the proper
trend in comparison with both STM \cite{Kliewer00S} and ARUPS
\cite{Reinert2001PRB} data. 

\begin{table}
\caption[]{Calculated and measured low-temperature lifetime widths
$\Gamma$ at band minimum in meV and mass enhancement factors $\lambda$
for the surface states of Cu(111) and Ag(111). $\Gamma_{ee}$ is the
calculated lifetime width due to electron-electron scattering
\cite{Kliewer00S},  $\Gamma_{ep}$ from electron-phonon scattering
(present calculation), $\Gamma_{STM}$ from STM measurements at T=4.6 K
\cite{Kliewer00S}, $\Gamma_{ARUPS}$ from ARUPS measurements
\cite{Reinert2001PRB}. }
\begin{ruledtabular}
\begin{tabular}{l c c c c c c }

& $\Gamma_{ep}$  & $\Gamma_{ee}$\footnote{This value for $\Gamma_{ee}$, is a lower limit, because
the used model does not take into account the
confinement of $d$ electrons \cite{Arantzazu}.} & $\Gamma_{ee} + \Gamma_{ep}$ & $\Gamma_{STM}$ & $\Gamma_{ARUPS}$ & $\lambda$ \\
Cu(111) & 6.6 & 14.0 & 20.0 & 24 & 23 $\pm$ 1.0 & 0.16 \\ 
Ag(111) & 3.6  & 2.0 & 5.5 & 6 &  6 $\pm$ 0.5 & 0.12 \\
\end{tabular}
\end{ruledtabular}
\label{table}
\end{table}

In summary we have presented results from a theoretical and
experimental study of the inherent phonon induced lifetime broadening
of surface states of the noble metal surfaces Cu(111) and
Ag(111). Taking into account the electron and phonon states, including
surface states is crucial for the understanding of experimental
findings in particular for states close to the Fermi level. The
Rayleigh surface phonons are shown to give an important contribution
to the phonon-induced lifetime broadening of surface states, in
particular for small binding energies. We have demonstrated that for noble
metal surface states, simple Thomas Fermi screening of Ashcroft pseudo
potentials gives very similar results as a more rigorous treatment of
the screening. We find excellent agreements with
experiments both with respect to the binding energy and the temperature 
dependence of the lifetime broadening. 

We acknowledge financial support from the Basque Government,
the Deutsche Forschungsgemeinschaft (SFB 277), and the Carl Tryggers
Foundation.

\begin{figure}
\caption{{\bf (a)}  The phonon dispersion from a 31 layer slab calculation in the $\bar{\Gamma}$-$\bar{M}$ direction of the S.B.Z.  
{\bf (b)} The Eliashberg function of the hole state in the $\bar{\Gamma}$ point (solid line) and 
the contribution from the Rayleigh mode to the Eliashberg function (dashed line).}
\label{Fig1}
\end{figure}

\begin{figure}
\caption{{\bf (a)} Lifetime broadening of the Cu(111) surface hole
  state as a function of binding energy, $\Gamma_{ee}$ + $\Gamma_{ep}$
  (solid line), $\Gamma_{ep}$ (dotted line), the Rayleigh mode
  contribution to $\Gamma_{ep}$ (dashed line) and photoemission data
  (diamonds). 
{\bf (b)} The same as in (a) for Ag(111).}
\label{Fig2}
\end{figure}

\begin{figure}
\caption{Lifetime broadening of the Cu(111) and Ag(111) surface hole
  states as a function of temperature (solid line), $\Gamma_{ee}$
  (dotted line) photoemission data (open circles). The inset shows
  the energy distribution curves (EDC) on Cu(111) for selected temperatures.} 
\label{Fig3}
\end{figure}

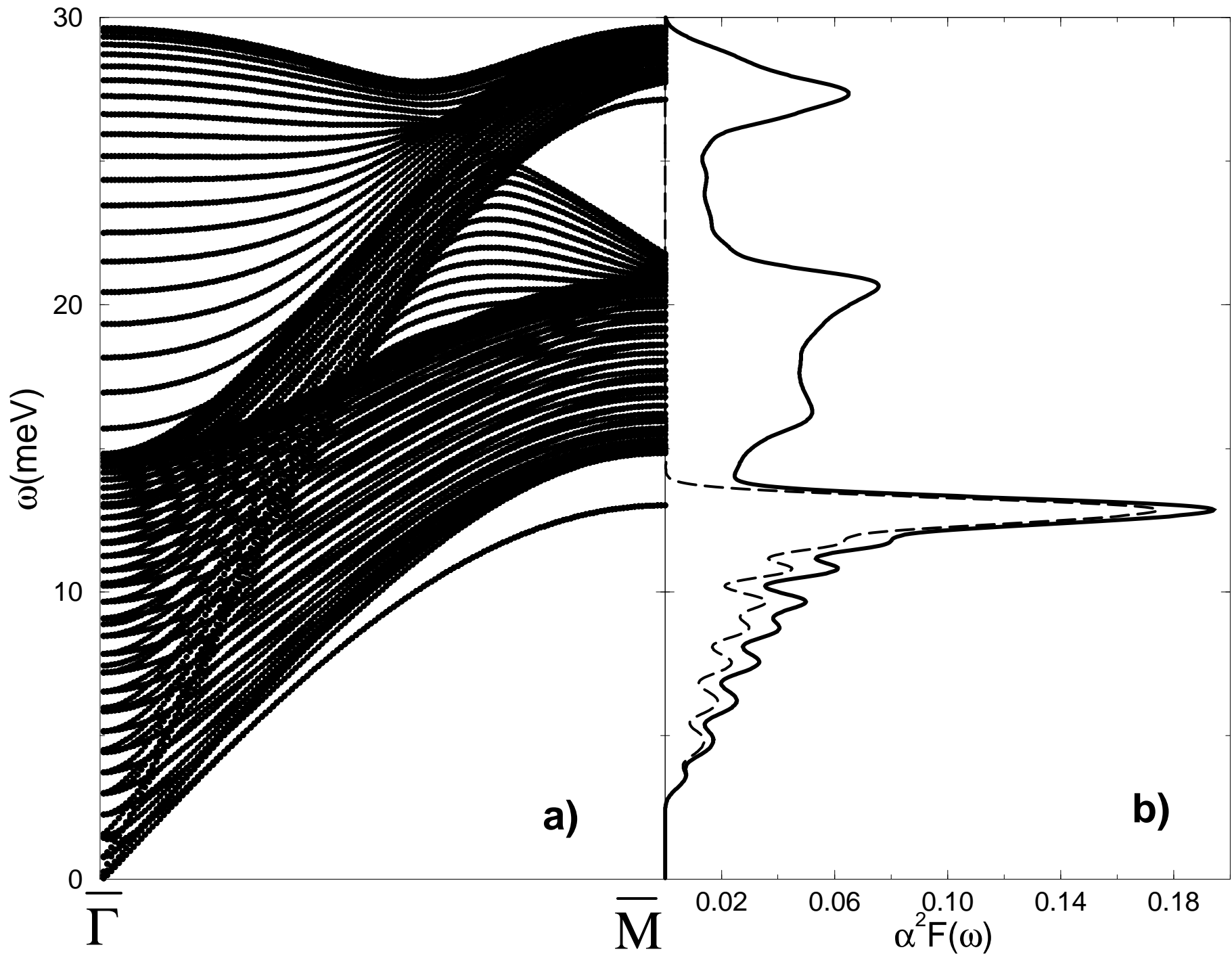
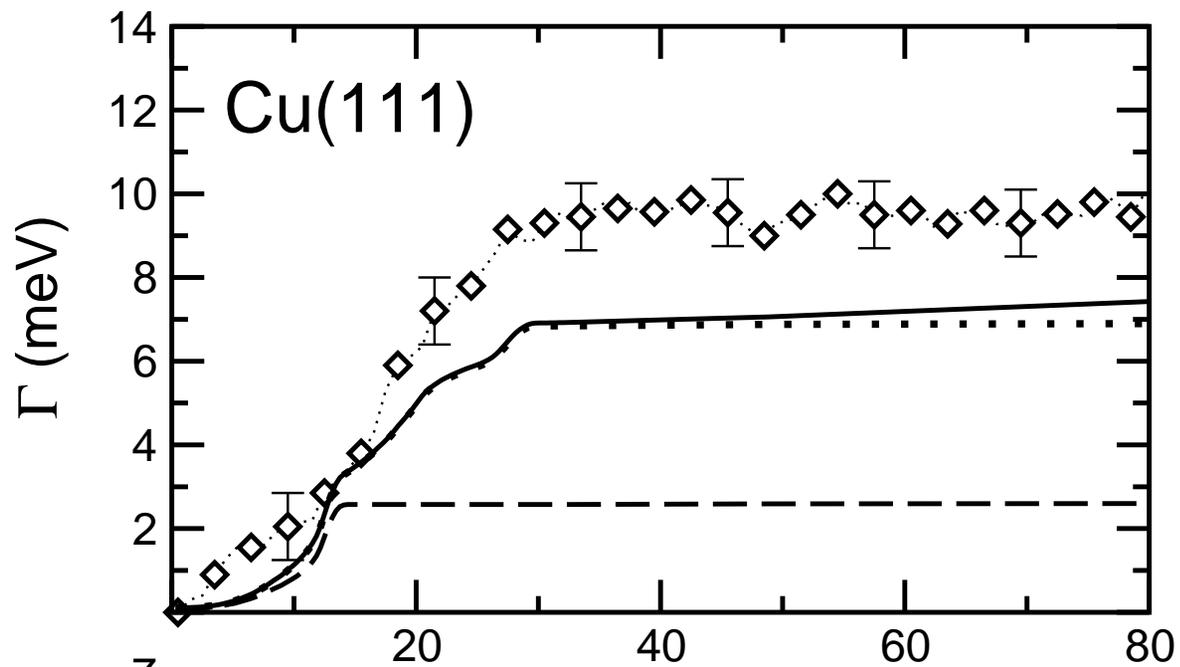
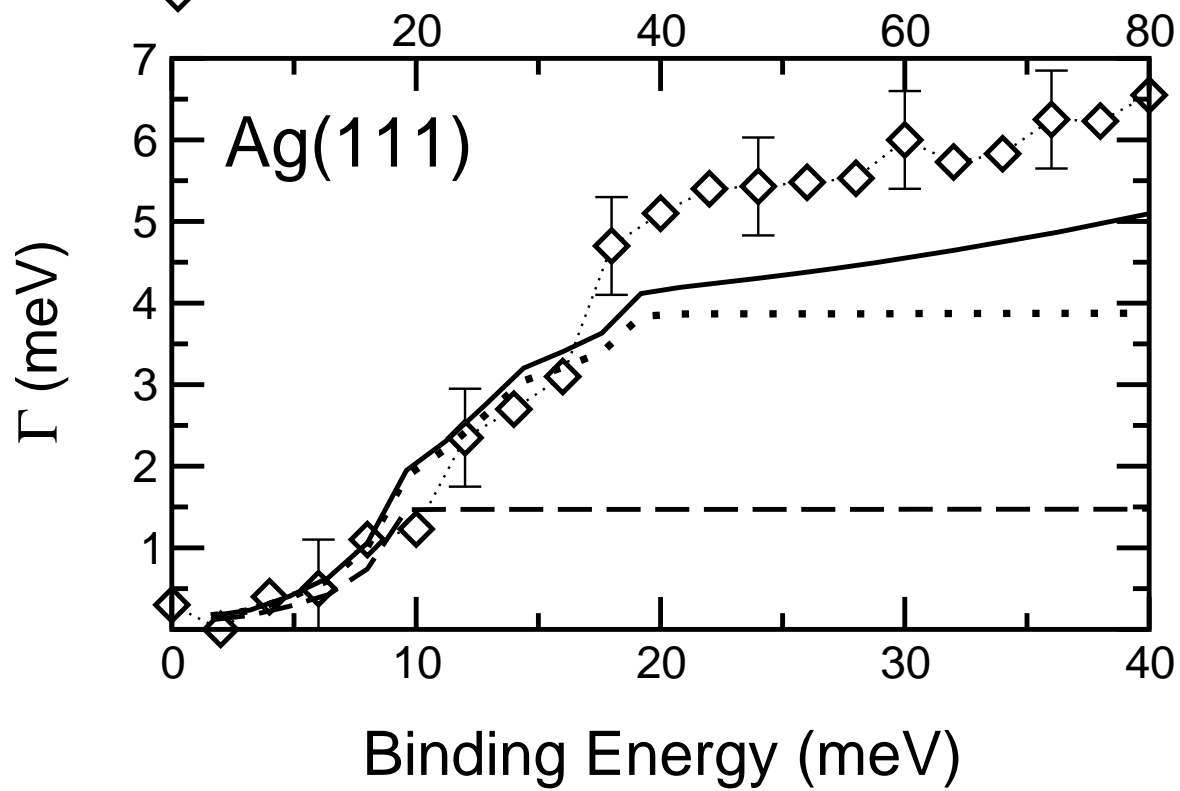
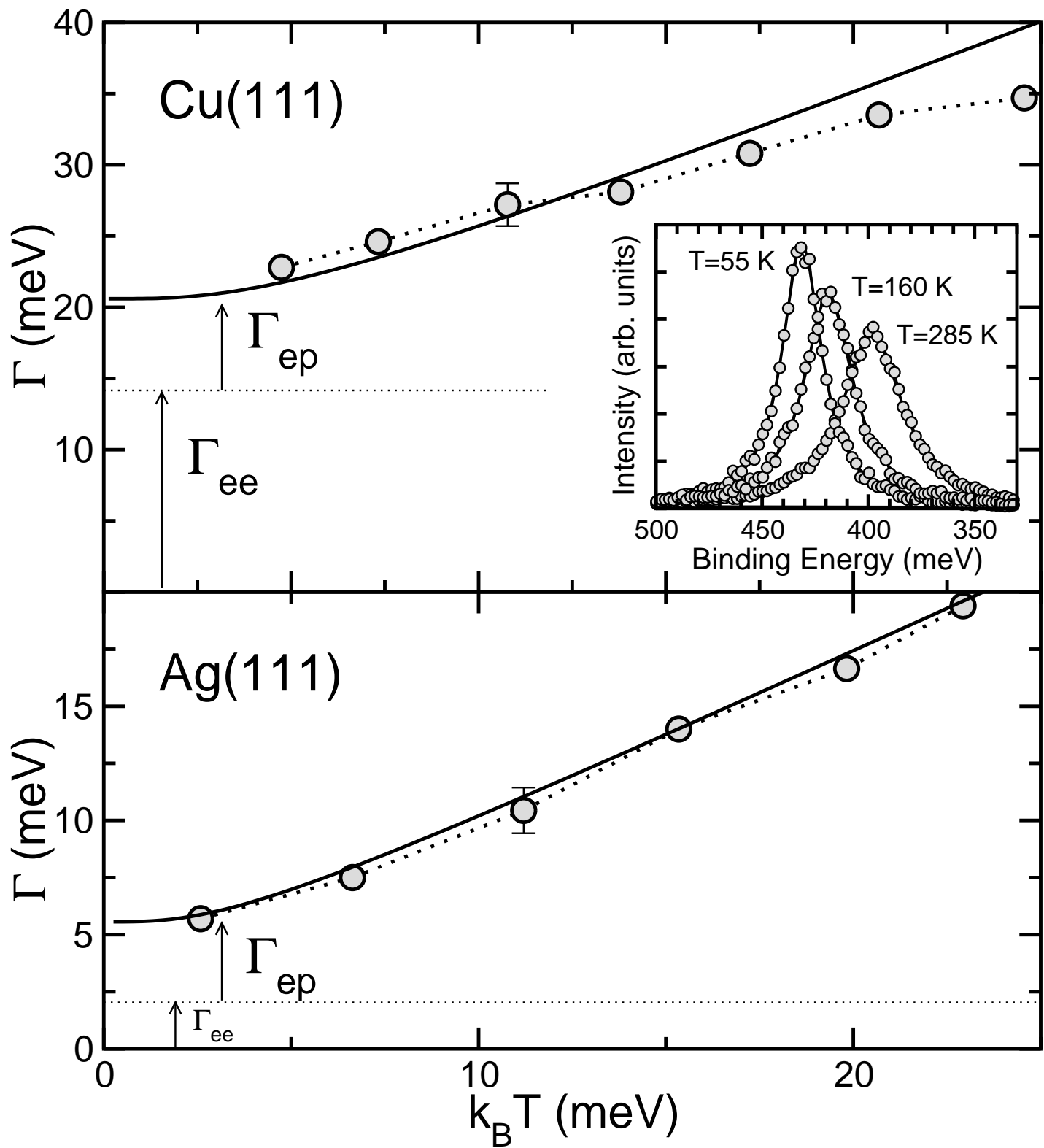
\end{document}